\documentclass[journal]{vgtc}                     % final (journal style)

\usepackage{amsmath}
\usepackage{amsfonts}
\usepackage{booktabs} % For formal tables
\usepackage{multirow}
\usepackage{titlesec}

\makeatletter
\DeclareRobustCommand\onedot{\futurelet\@let@token\@onedot}
\def\@onedot{\ifx\@let@token.\else.\null\fi\xspace}
\def\eg{e.g\onedot}

\makeatother

\usepackage[normalem]{ulem}

\title{ImmerseGen: Agent-Guided Immersive World Generation with Alpha-Textured Proxies}

\author{%
  \authororcid{Jinyan Yuan}{0009-0001-3139-7616},
  \authororcid{Bangbang Yang}{0000-0001-7604-5553},
  \authororcid{Keke Wang}{0009-0006-1854-2539},
  \authororcid{Panwang Pan}{0000-0001-8631-012X},
  \authororcid{Lin Ma}{0009-0005-2568-2735},\\
  \authororcid{Xuehai Zhang}{0009-0000-9972-2916},
  \authororcid{Xiao Liu}{0000-0002-6073-030X}, 
  \authororcid{Zhaopeng Cui}{0000-0002-7130-439X},
  and 
  \authororcid{Yuewen Ma}{0009-0003-7734-7053}
}

\authorfooter{
  \item
  	Jinyan Yuan, Bangbang Yang, Keke Wang, Panwang Pan, Lin Ma, Xuehai Zhang, Xiao Liu and Yuewen Ma (corresponding author) are with Bytedance. 
    E-mail: \{yuanjinyan | yangbangbang | wangkeke.1014 | panpanwang | malin.ml | xuehai.zhang | liuxiao.ai | mayuewen\}@bytedance.com 
  \item
  	Zhaopeng Cui is with Zhejiang University. E-mail: zhpcui@zju.edu.cn
  \item
    Jinyan Yuan and Bangbang Yang contributed equally to this work.
}

\abstract{Automating immersive VR scene creation remains a primary research challenge. Existing methods typically rely on complex geometry with post-simplification, resulting in inefficient pipelines or limited realism.
In this paper, we introduce ImmerseGen, a novel agent-guided framework for compact and photorealistic world generation that decouples realism from exhaustive geometric modeling.
ImmerseGen represents scenes as hierarchical compositions of lightweight geometric proxies with synthesized RGBA textures, facilitating real-time rendering on mobile VR headsets.
We propose terrain‑conditioned texturing for base world generation,  combined with context-aware texturing for scenery, to produce diverse and visually coherent worlds.
VLM-based agents employ semantic grid-based analysis for precise asset placement and enrich scenes with multimodal enhancements such as visual dynamics and ambient sound.
Experiments and real-time VR applications demonstrate that ImmerseGen achieves superior photorealism, spatial coherence, and rendering efficiency compared to existing methods.}


\teaser{
  \centering
  \includegraphics[width=0.93\linewidth]{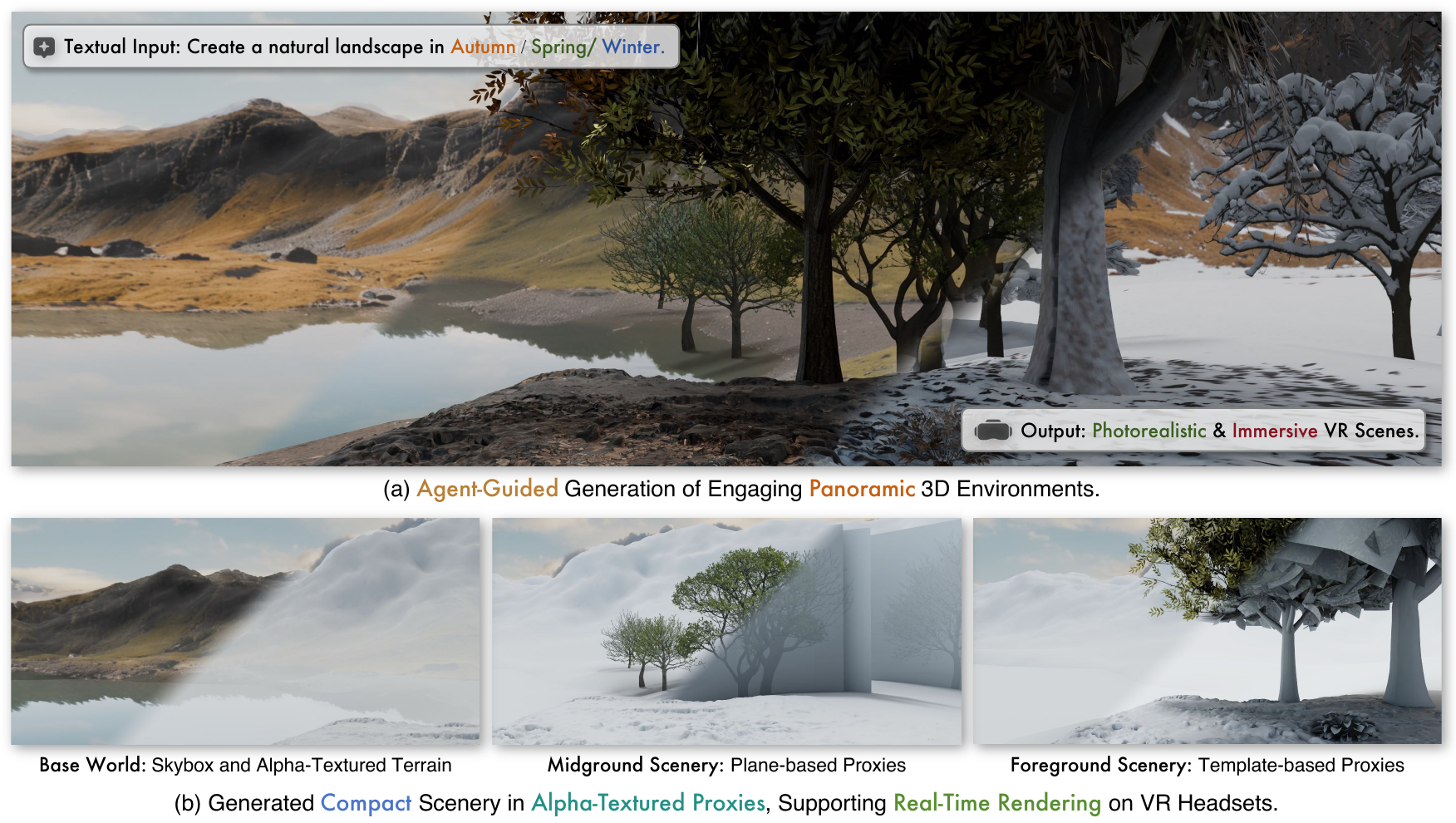}
    \caption{Tailored for immersive VR experiences, ImmerseGen generates photorealistic worlds from text prompts by synthesizing compact alpha-textured proxies through agent-guided design and arrangement, obviating the need for complex assets while ensuring diversity.
    }
  \label{fig:teaser}
}

\graphicspath{{figs/}{figures/}{pictures/}{images/}{./}} % where to search for the images

\usepackage{tabu}                      % only used for the table example
\usepackage{booktabs}                  % only used for the table example
\usepackage{lipsum}                    % used to generate placeholder text
\usepackage{mwe}                       % used to generate placeholder figures

\usepackage{mathptmx}                  % use matching math font

\begin{document}

%%%%%%%%%%%%%%%%%%%%%%%%%%%%%%%%%%%%%%%%%%%%%%%%%%%%%%%%%%%%%%%%
%%%%%%%%%%%%%%%%%%%%%% START OF THE PAPER %%%%%%%%%%%%%%%%%%%%%%
%%%%%%%%%%%%%%%%%%%%%%%%%%%%%%%%%%%%%%%%%%%%%%%%%%%%%%%%%%%%%%%%

\firstsection{Introduction}

\maketitle

Humans have an innate desire to create and inhabit personalized worlds, whether it’s children building sandcastles or artists designing landscapes.
This creative drive extends to digital spaces, especially in VR/XR applications, where users expect to be immersed in custom environments with panoramic views, high-fidelity visuals, and real-time interactions. 
However, building such immersive 3D scenes remains challenging.
Handcrafted 3D modeling requires specialized skills and considerable effort,
while recent generative methods like object-compositional generation~\cite{huang2024midi, yao2025cast,engstler2025syncity}, LLM-powered modeling tools~\cite{BlenderMCP} and frameworks\cite{liu2025worldcraft,scenethesis, scenex}, and approximating through 3D Gaussians~\cite{dreamscene360, wonderworld, layerpano3d} often struggle to balance photorealism with computational efficiency.
These approaches prioritize fully detailed geometry or massive Gaussians to achieve realism, but often result in overly complex scene representations that hinder real-time performance on VR headsets, or require handcrafted and time-consuming decimation and compression to make them usable.
This raises a fundamental issue regarding the necessity of exhaustive 3D modeling for immersive VR. We argue that starting with complex geometry is not a prerequisite, especially when considering the limited explorable areas and finite computational budgets.

In this paper, we propose ImmerseGen, a novel agent-guided framework that models immersive scenes as hierarchical compositions of lightweight RGBA-textured geometric proxies, including simplified terrain meshes and alpha-textured billboard meshes.

The formulation offers several important advantages:

\textbf{1)} 
Rather than modeling the scene with complex geometry and then simplifying it, our approach bypasses this process by generating photorealistic texture directly on lightweight geometric proxies leveraging SOTA image generators, alleviating reliance on detailed asset creation and preserving the texture quality without artifacts introduced in decimation or Gaussian approximations.

\textbf{2)}
Such modeling paradigm enables agents to flexibly guide generative models in synthesizing coherent, context-aware textures that integrate seamlessly with the panoramic world;

\textbf{3)} It delivers VR-ready scene representations that allow real-time rendering at smooth frame rates.

To establish this hierarchical paradigm, ImmerseGen first creates the base layer world, which employs a terrain-conditioned RGBA texturing scheme on a simplified terrain mesh with user-centric UV mapping.
More specifically, it employs a user-centric texturing and mapping scheme that synthesizes and allocates higher texture resolution based on central camera origin, prioritizing the primary viewing area, rather than uniformly covering the entire scene with limited quality~\cite{engstler2025syncity, infinigen2023infinite}.
Then, ImmerseGen automatically enriches the environment with generative scenery assets, which are clearly separated into distinct depth levels.
Midground assets, such as distant trees or vegetation, are efficiently created using planar billboard textures, while foreground assets, closer to the user, are generated with alpha-textured cards placed over retrieved low-poly 3D template meshes.
This mechanism smartly allocates representation detail, maintaining both visual fidelity and rendering efficiency at every scale.

While RGBA-textured proxies simplify asset modeling, assembling coherent 3D scenes still requires manual adjustment and expert knowledge.
To simplify this process, we develop a Visual-Language Models (VLMs)-based agentic system that interprets user text prompts into immersive environments.
However, directly using VLMs often faces challenges in spatial understanding that hinder layout accuracy.
To address this, we introduce a grid-based semantic analysis strategy, enhancing the spatial comprehension with coarse-to-fine visual prompt and raycasting-based validation, thus mitigating placement errors and inconsistencies existing in na\"ive VLMs.
Moreover, ImmerseGen enriches the immersive experience by incorporating modular dynamics (e.g., flowing water, drifting clouds) and ambient audio (e.g., wind, birdcalls), delivering a fully multisensory environment.

In summary, our contributions are as follows:

\textbf{1)} We propose ImmerseGen, a novel agent-guided 3D environment generation framework that uses simplified geometric proxies with alpha-textured meshes to produce compact, photorealistic worlds ready for real-time mobile VR rendering.

\textbf{2)} We propose a novel RGBA texturing paradigm that first synthesizes 8K terrain textures using a geometry-conditioned panorama generator via user-centric mapping, and then directly generates alpha-textured proxy assets, avoiding fidelity loss inherent in mesh decimation.

\textbf{3)} To automate scene creation from user prompts, we introduce VLM-based modeling agents equipped with a novel grid-based semantic analysis, enabling 3D spatial reasoning from 2D observations and ensuring accurate asset placement. ImmerseGen further enhances immersion with dynamic effects and ambient audio for a multisensory experience.

\textbf{4)} Experiments on multiple scene-generation scenarios and live mobile VR applications show that ImmerseGen outperforms previous methods in visual quality, realism, spatial coherence, and rendering efficiency for immersive real-time VR experiences.

\begin{figure}[t!]
    \centering
    \includegraphics[width=1.0\linewidth, trim={0 0 0 0}, clip]{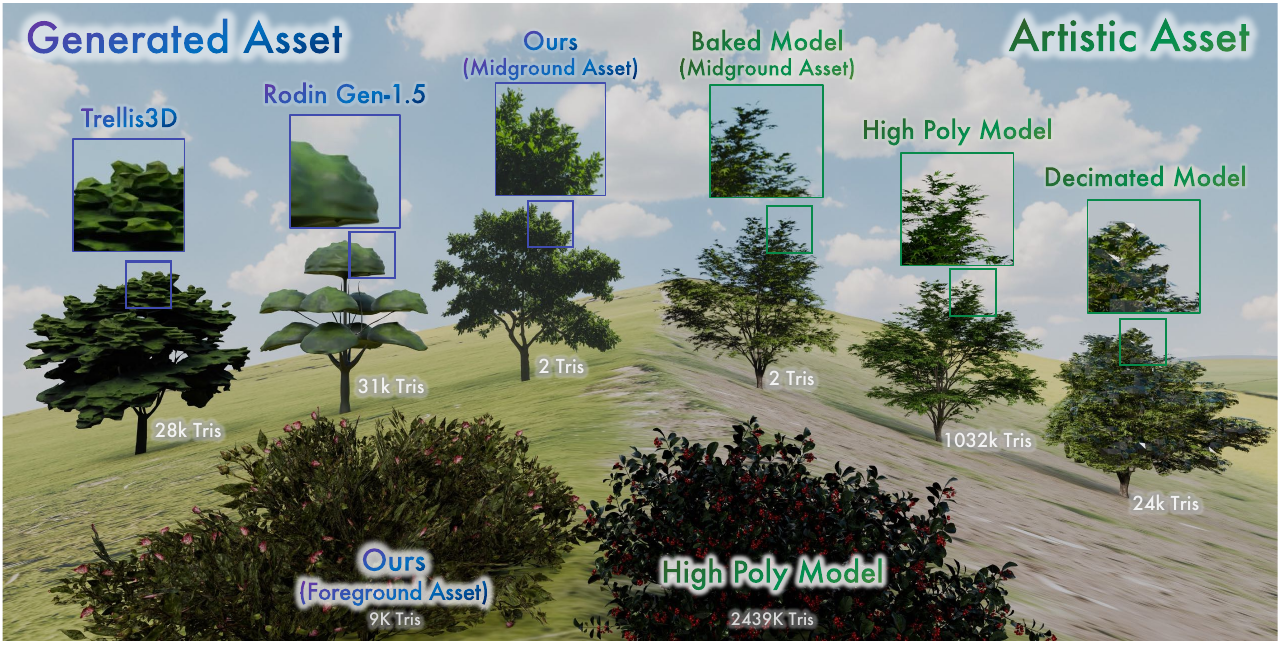}
    \caption{
    \textbf{Asset comparison from different sources.}
    We compare assets created by learning-based generative methods (blue labels), artists (green labels), and ours.
    Our generative RGBA-textured proxy assets achieve better visual details than existing models~\cite{clay,xiang2024structured} with fewer triangles, delivering photorealistic appearance comparable to artist-created high-poly or baked assets.
    }
    \label{fig:method_repr_compare}
\end{figure}

\begin{figure*}[t!]
    \centering
    \includegraphics[width=1.0\linewidth, trim={0 0 0 0}, clip]{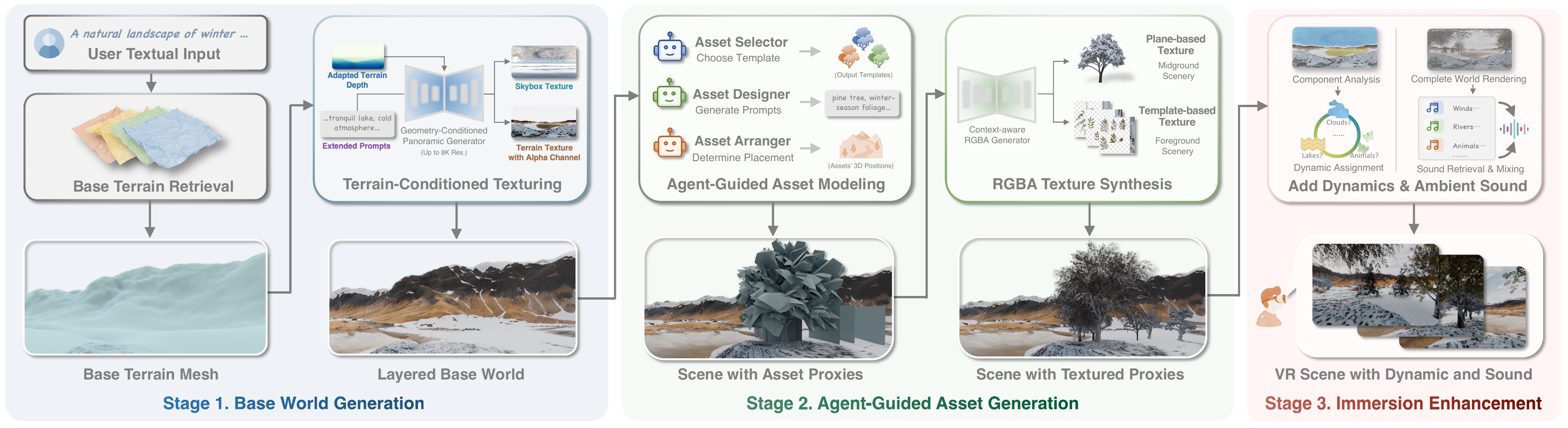}
    \caption{
    \textbf{Overview.}
    Given a user’s text input, the agent first retrieves a base terrain. Conditioned on the terrain depth and an extended prompt, panoramic textures for the terrain and sky are generated to form a layered base world. Next, VLM-based asset agents enrich the scene by selecting asset proxies as foreground or midground scenery, designing detailed asset prompts, and determining optimal asset placement. Each asset is instantiated via RGBA texture synthesis. Finally, the agent incorporates dynamic visual effects and synthesized ambient sound, producing a lightweight and photorealistic world.
    }
    \label{fig:framework}
\end{figure*}

\section{Related works}

\subsection{Agentic Scene Generation} 
Early efforts in procedural content generation (PCG) for immersive environments primarily rely on rule-based systems~\cite{parish2001procedural, lipp2011interactive, zhang2019procedural, gasch2022procedural}, where spatial relationships and asset placements are meticulously defined through handcrafted rules. Infinigen~\cite{infinigen} advances this process by leveraging Blender scripts to orchestrate multiple procedural generators, enabling the creation of larger and more complex scenes. However, PCG methods inherently limit adaptability to novel scenarios and user-driven instructions.
The advent of LLMs and VLMs introduces a paradigm shift in scene generation, enabling more intuitive, instruction-based workflows. Recent methods like BlenderMCP~\cite{BlenderMCP} increasingly harness the capabilities of LLMs to automate the generation process, employing function-calling agents to interpret text prompts~\cite{SceneTeller,Holodeck,GALA3D}, design scene layouts~\cite{LayoutVLM,lin2024instructscene}, and populate environments~\cite{scenex,scenethesis} with assets retrieved from pre-built libraries~\cite{BlenderMCP,3d_gpt,scenex,scenecraft,liu2025worldcraft,liu2024controllable}. These systems demonstrate significant potential in generating diverse, large-scale scenes from high-level descriptions, streamlining the content creation pipeline.
However, existing LLM/VLM-based approaches rely heavily on asset libraries, often requiring a trade-off between quality and efficiency. Moreover, the precision of VLM-guided asset placement often proves insufficient in complex scenarios. In contrast, ImmerseGen addresses these limitations by introducing lightweight proxy assets and semantic grid-based arrangement by agents, enabling the creation of compact, photorealistic worlds.

\subsection{Learning-based Generation}
Recently, learning-based generation methods have shown promising results in creating 2D and 3D content~\cite{stable_diffusion,clay,triplane_gaussian,lrm}. However, unlike 3D object generation that benefits from diverse object datasets~\cite{yu2023mvimgnet,deitke2023objaversexl} for model training, 3D scene generation still faces challenges~\cite{blockfusion,lt3sd,text2room,sketch2scene,huang2024midi} due to the lack of comprehensive scene-level data and unified representations.
Early methods either learned a generative neural field with GAN~\cite{scenedreamer,gancraft,infinicity,citydreamer} or 2D diffusion priors~\cite{text2nerf,set_the_scene,scenewiz3d}, but failed to produce detailed appearance.
Recently, other lines of work tend to generate images and lift them to 3D space through depth prediction, combined with outpainting techniques to expand the scene~\cite{luciddreamer,wonderjourney,wonderworld,scenescape}.
However, these methods typically produce incomplete 3D worlds (\eg, missing 360-degree views or geometry under the feet), thus failing to meet the demands of immersive VR applications.
To create a complete surrounding world, some methods lift the generated panoramic images~\cite{panfusion,stitch_diffusion} to 3D space with depth estimation and inpainting~\cite{dreamscene360,holodreamer,li20244k4dgenpanoramic4dgeneration,layerpano3d}, but still faces challenges in producing spatially coherent worlds due to the inconsistency of novel view inpainting.
More recent approaches utilize video models for 3D scene creation~\cite{cat3d,dimension_x,wonderland,splatflow}, which either suffer from blurry backgrounds or fail to guarantee fully explorable 360-degree environments.
Additionally, these methods often produce a large number of point clouds or 3D Gaussians for scene representation, making it challenging to achieve high-quality rendering while maintaining reasonable computational costs.

\subsection{Traditional Asset Creation}
Conventional asset creation pipelines typically follow a two-stage process: detailed geometric modeling followed by texture mapping.
This modeling-first paradigm is prevalent in CG content production where artists craft complex meshes and apply high-resolution textures to achieve visual realism.
However, when deploying such assets in real-time rendering applications like VR or games, these models require simplification via decimation techniques, such as mesh simplification~\cite{li2018optcuts,liu2017seamless}, billboard generation~\cite{decoret2003billboard,kratt2014adaptive}, or level-of-detail (LOD) hierarchies~\cite{huang2025arcpro,zhang2024architectural}, along with baked textures. For natural scenes, many works on terrain generation and vegetation modeling~\cite{li2021learning,lee2023latent} have been proposed, yet they often lack diversity and realism.
While effective, this conventional workflow incurs significant manual effort or computational cost, as it first generates  excessively detailed primitives only to later reduce their complexity for efficiency.
In contrast, ImmerseGen eliminates the need for post-hoc simplification by directly synthesizing alpha-textured assets tailored for efficient rendering, enabling photorealistic scene generation optimized for immersive applications.

\section{Method}

We introduce ImmerseGen, an agent-guided framework for generating immersive 3D scenes from textual prompts.
As shown in Fig.~\ref{fig:framework}, we construct the scene hierarchically from base terrain guided by VLM-based agents.
First, we generate a layered base world via terrain-conditioned texturing, where panoramic sky and RGBA terrain textures are synthesized upon a retrieved terrain mesh (Sec.~\ref{ssec:method_base_world}).
Next, we enrich the scene through placing lightweight asset proxies by agents with semantic grid-based analysis. The selected assets are then instantiated using a context-aware RGBA texture synthesis scheme (Sec.~\ref{ssec:method_asset_gen}).
Finally, we augment the scene with dynamic effects guided by agents, such as flowing water and ambient sound, delivering a multisensory experience (Sec.~\ref{ssec:method_immersion_enhance}).

\begin{figure}[t!]
    \centering
    \includegraphics[width=1.0\linewidth, trim={0 0 0 0}, clip]{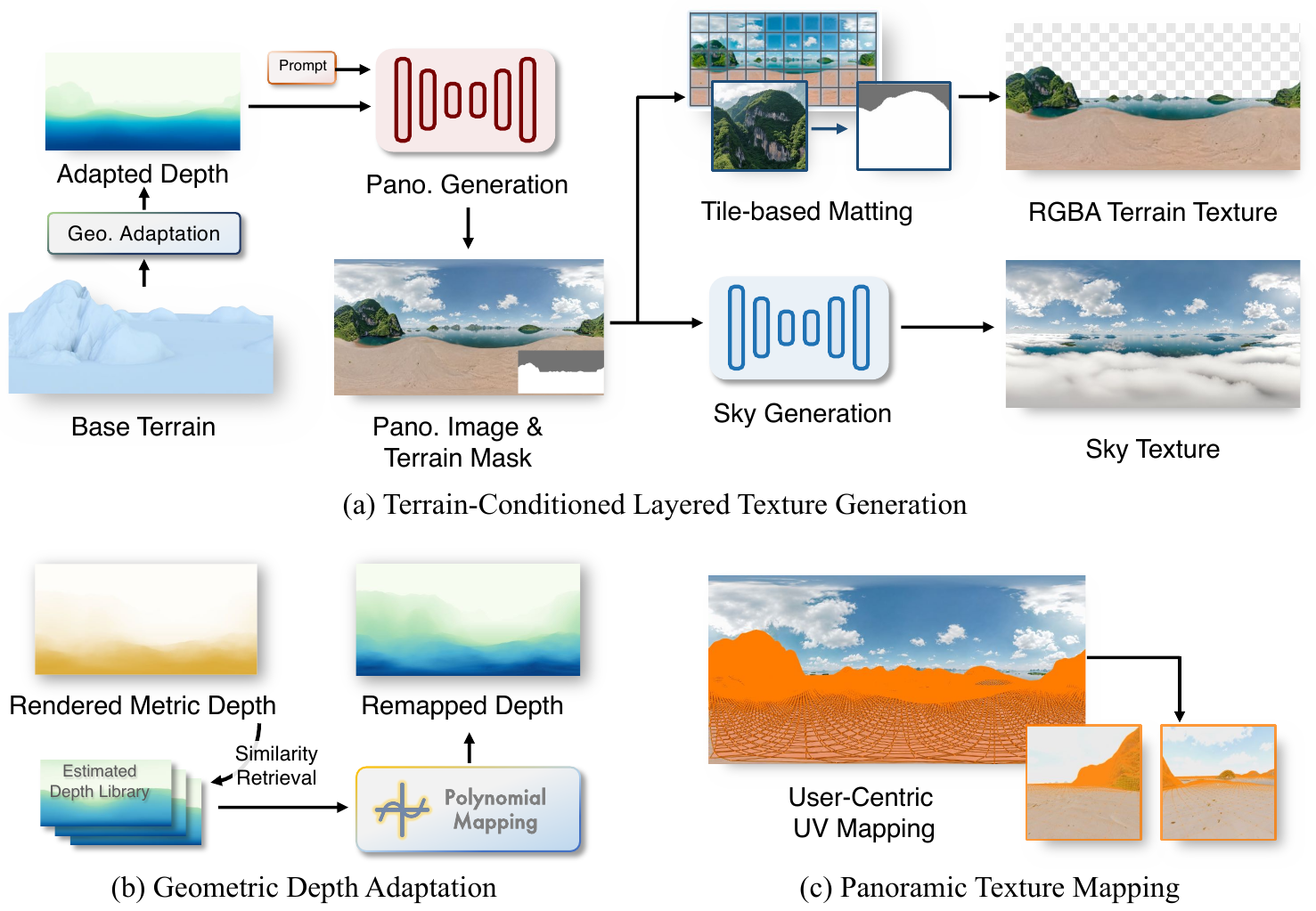}
    \caption{
    \textbf{Workflow of base world generation.}
    Panoramic textures for terrain mesh and sky are generated for the base world. To tame the diffusion model for terrain texturing, we propose geometric adaptation (b) for depth control and user-centric texture mapping (c).
    }
    \label{fig:method_terrain}
\end{figure}

\subsection{Base World Generation}
\label{ssec:method_base_world}

\paragraph{From textual prompts to base terrain.}
Given a user's textual prompt describing the world, a suitable base terrain mesh is first retrieved from a pre-generated template library.
These templates are created using procedural content generation tools, followed by post-processing steps including remeshing, visibility culling, and captioning to support effective retrieval.
We deploy an LLM agent that retrieves a suitable terrain template and enhances prompts with imaginative and contextually relevant details from user's textual input, to improve scene diversity and ensure coherence between terrain characteristics and the input prompt. 
Since visual diversity is primarily introduced through subsequent generative texturing, this strategy strikes a practical balance between efficiency and variety. 

\paragraph{Terrain-conditioned texturing.}
As demonstrated in Fig~\ref{fig:method_terrain} (a), given a base terrain mesh and text prompts, we first generate panoramic sky texture and alpha ground textures upon the mesh.
To support terrain texture synthesis in equirectangular projection (ERP), we adopt a two-stage training pipeline.
We first train a panoramic diffusion model built upon Stable Diffusion XL\cite{podellsdxl} on ERP data conditioned on textual prompts~\cite{stable_diffusion}.
Then, we extend this model by training a depth-conditioned ControlNet~\cite{zhang2023adding}, which takes as input a panoramic depth map $\mathbf{D}_\mathcal{M}$  estimated from a neural depth estimator~\cite{depth_anything_v2}.
During inference, we combine both modules to generate a panoramic texture $\mathbf{I}_t$ that aligns with the terrain mesh $\mathcal{M}$, formulated as:
\begin{equation}
    \mathbf{I}_t = \mathcal{U}(\mathcal{G}(\mathbf{D}_\mathcal{M}; \mathcal{C}_\text{Global},\mathcal{C}_\text{Region})),
\end{equation}
where $\mathbf{D}_\mathcal{M}$ is the conditioning panoramic depth map rendered from the terrain mesh, $\mathcal{G}$ is the conditional diffusion model, $\mathcal{C}_\text{Global}$ is the text prompt for global geographic description, $\mathcal{C}_\text{Region}$ is the optional regional prompts for generating designated geographic features (such as water body, see supp. Sec.~1.1 ), and $\mathcal{U}$ is the conditioned upscaling model that produces 8K textures to enhance details utilizing a tile-based generation approach inspired by~\cite{bar2023multidiffusion}.  

To separate the terrain texture and sky texture while maintaining high resolution, we perform tile-based matting and sky outpainting on the panorama, which yields an 8K fine-grained alpha matte and pure sky texture guided on the terrain mask.
This detailed alpha matte allows low-poly terrain meshes to exhibit highly detailed landscape silhouettes (e.g., trees and houses against the sky).

\begin{figure}[t!]
    \centering
    \includegraphics[width=1.0\linewidth, trim={0 0 0 0}, clip]{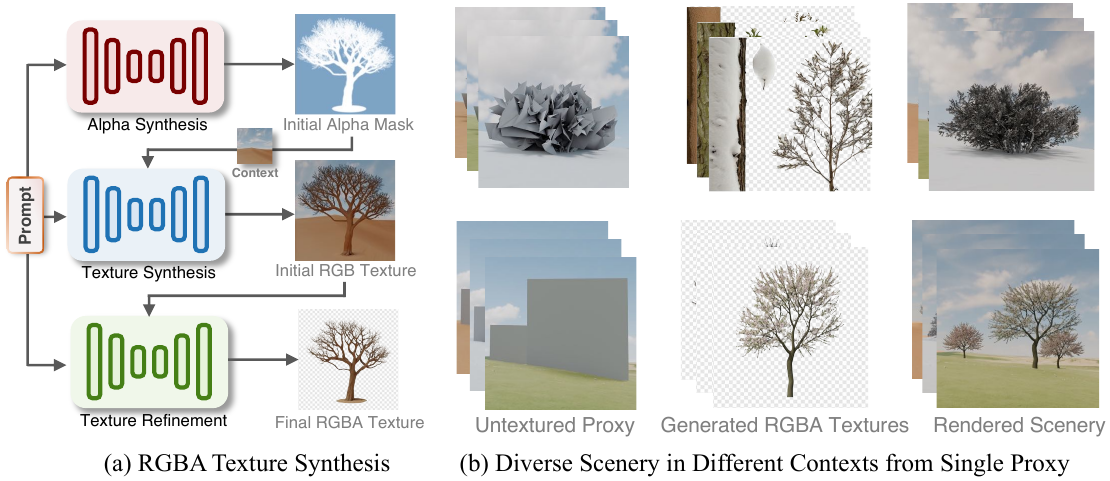}
    \caption{
    The proposed \textbf{context-aware texture synthesis} (a) generates diverse, contextually coherent RGBA textures directly on lightweight proxies for both foreground and midground scenery (b).
    }
    \label{fig:method_lod}
\end{figure}

\paragraph{Depth control with geometric adaptation.}

While it is technically plausible to apply conditional diffusion for mesh texturing, we find it non-trivial to produce 3D-coherent textures that align well with the terrain and meet immersive standards, i.e., degraded quality as shown in Fig.~\ref{fig:exp_ablation_terrain_cond}.
This difficulty arises primarily from the domain gap between the estimated relative depth for ControlNet training and rendered metric depth maps for inference-time conditioning. 
To tackle this issue, we propose a geometric adaptation scheme that remaps the rendered metric depth to better match the domain of training-time estimated depth.
Specifically, we retrieve the most similar depth map $\mathbf{D}_{\text{Retrieve}}$ from a sampled training set $\mathcal{L}$ using cosine similarity, and apply a polynomial remapping function:
\begin{equation}
    \hat{\mathbf{D}}_\mathcal{M} = \mathcal{P}(\mathbf{D}_\mathcal{M}; \mathbf{D}_{\text{Retrieve}}),
\end{equation}
where $\hat{\mathbf{D}}_\mathcal{M}$ is the remapped depth, and $\mathcal{P}$ is a third-degree polynomial mapping function.
Practically, we downsample both $\mathbf{D}_\mathcal{M}$ and $\mathbf{D}_{\text{Retrieve}}$ to $32 \times 16$ resolution to estimate the polynomial coefficients, which are then applied to the full-resolution depth map $\mathbf{D}_\mathcal{M}$. By conditioning on the remapped depth, the panoramic diffusion model generates textures that are well aligned with the terrain.

\paragraph{Terrain texture mapping.}
To efficiently texture the terrain with the generated panoramic texture while preserving visual fidelity, we precompute user-centric panoramic UV coordinates for the terrain mesh, as illustrated in Fig.~\ref{fig:method_terrain} (c).
Thus, the texture can be directly sampled during the rendering without back-projection or baking procedures.
Specifically, the UV coordinate for each mesh vertex can be calculated by transforming the coordinates from object space to camera space. 
Given a mesh vertex position in camera space $\mathbf{p} = (x, y, z)^\top$, the corresponding UV coordinate $\mathbf{u} = (u, v)^\top$ on the panoramic texture $\mathbf{I}_t$ can be calculated as:
\begin{equation}
\mathbf{u} = \left(\frac{1}{2\pi}\arctan(\frac{x}{-z}) + \frac{1}{2}, \frac{1}{\pi}\arcsin(\frac{y}{\|\mathbf{p}\|}) + \frac{1}{2}\right)^\top,
\end{equation}
where $\|\mathbf{p}\|$ denotes the L2-norm of the vertex position.
To prevent texture stretching at horizontal seams, we detect UVs crossing the panoramic boundary and offset them for correct wrapping, then set the texture wrapping mode to \textit{repeat} for seamless interpolation of panoramic texture sampling.

To further improve visual fidelity around the user's viewpoint, particularly in the polar region where the ERP exhibits stretching, we first adopt an ERP-to-cubemap refinement scheme, using an image-to-image diffusion method~\cite{meng2021sdedit} to repaint the bottom area.
Then, we partition the mesh by cropping its bottom area and then reassign UV coordinates of this mesh to directly sample textures from the bottom map. 
Additionally, to achieve better geometric realism, we incorporate a displacement map obtained from an adapted depth estimation model~\cite{depth_anything_v2} (see supp. Sec.~1.5.).

\subsection{Agent-Guided Asset Generation}
\label{ssec:method_asset_gen}

To enrich the base world with photorealistic scenery, we then add more generative 3D assets (such as vegetation) to the scene.
Unlike prior methods that rely on complex modeling pipelines~\cite{decoret2003billboard} or off-the-shelf asset retrieval, our framework dynamically generates unique, alpha-textured asset proxies from coarse templates using generative texture synthesis, thus simplifying asset creation and enabling more flexible agent-driven design.

\paragraph{Defining proxies by distance.}
We employ distinct proxy types based on the distance between the user and the asset to balance rendering quality and performance, which delivers a realistic appearance comparable to the artists' baked models while alleviating the cost of baking or decimation.
As demonstrated in Fig.~\ref{fig:teaser} (b) and Fig.~\ref{fig:method_repr_compare}, 
for midground objects, since users cannot perceive detailed depth changes of object surfaces, we synthesize RGBA textures on distant planar mesh (see Fig.~\ref{fig:method_lod} (c), a.k.a. billboard texture).
For foreground objects that require depth perception, we generate alpha textures for each group of shared materials in a template mesh with alpha cards (such as tree leaves and trunks, see Fig.~\ref{fig:method_lod} (b)).

\begin{figure}[t!]
    \centering
    \includegraphics[width=1.0\linewidth, trim={0 0 0 0}, clip]{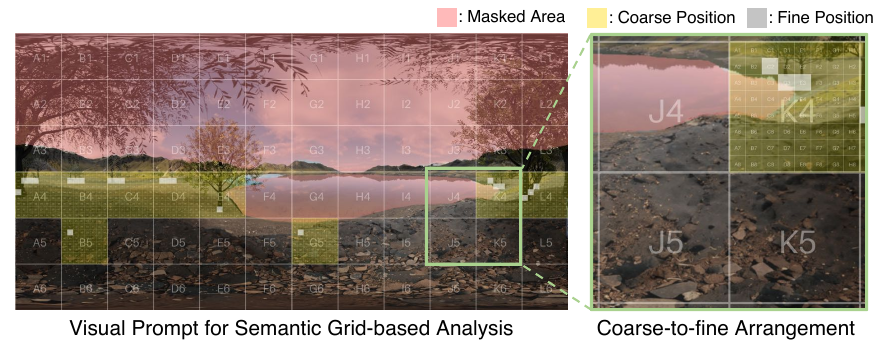}
    \caption{
    The proposed \textbf{semantic grid-based analysis} overlays a labeled grid and masks unsuitable regions as visual prompts. This enables the VLM agent to progressively select grid cells in a coarse-to-fine manner, improving the accuracy and semantic coherence of asset arrangement.}
    \label{fig:method_agent}
\end{figure}

\paragraph{Asset selection and designing.}
To create diverse and contextually coherent scenery assets, we develop VLM-based agents to guide the asset design pipeline.
First, the \textbf{asset selector} analyzes the rendered base world image and user's textual description to retrieve suitable foreground asset templates from an offline-generated library indexed by description, e.g., pine trees for mountainous regions or bushes for arid deserts.
Next, the \textbf{asset designer} crafts detailed textual prompts to guide generative models in synthesizing these scenery assets.
In practice, the designer examines both the generated base-world image and selected texture templates, and produces detailed descriptions for each scenery asset (such as categories, season, styles, etc.).

\paragraph{Asset arrangement with semantic grid-based analysis.}
To ensure that generative assets are placed in semantically appropriate and visually plausible locations, we introduce an \textbf{asset arranger} that analyzes the base world image to produce 2D position candidates, which are then back-projected to determine 3D positions through raycasting and validation.
One primary challenge for the asset arranger is to generate reasonable 3D placements based solely on image-based observation.
A na\"ive approach is to let the agent directly output the coordinate, which generally results in inaccurate positions and meaningless layout (see Sec.~\ref{ssec:exp_layout}) due to the limited spatial understanding ability of existing models~\cite{yang2024thinking}.
To address this, we propose a semantic grid-based position proposal scheme, which significantly improves the asset arrangement quality.
As shown in Fig.~\ref{fig:method_agent}, we overlay the base world image with a labeled grid and mask out unsuitable regions (e.g., water, sky), forming a structured visual prompt for the VLM agent.
The agent first selects coarse grid cells given this visual prompt.
Then, for finer placement, each selected cell is zoomed in and subdivided into sub-grids, from which the agent will select a more precise sub-cell.
The final positions are determined by randomly selecting a point within the sub-cell.

\paragraph{Context-aware RGBA texture synthesis.}
Once the agents have determined the per-asset placement and textual descriptions, we proceed to instantiate each asset by synthesizing its RGBA texture in context with the base world.
To facilitate seamless integration, we propose a context-aware cascaded RGBA texture synthesis model conditioned on base world background textures, which is inspired by the layered diffusion model~\cite{layerdiffuse}.

Given a scenery prompt $\mathcal{C}_s$, the alpha synthesis module $\mathcal{G}_a$ first generates an alpha mask $\mathbf{M}_c = \mathcal{G}_a ( \mathcal{C}_s ) \in \mathbb{R}^{H \times W} $, serving as a sketch for subsequent texturing. 
To incorporate contextual information from the base world, the RGB base texture reference $\mathbf{I}_b \in \mathbb{R}^{H \times W \times 3}$ is injected into an initially empty RGBA canvas through alpha blending guided by $\mathbf{M}_c$.
Then the texture synthesis module $\mathcal{G}_i$ generates an initial scenery texture from the alpha-blended reference with the alpha mask $\mathbf{M}_c$.
Note that the generated texture usually produces boundaries that is not perfectly aligned with the given alpha mask. Thus, the alpha channel of the initial texture is further refined through a diffusion-based refinement module $\mathcal{R}$.
The full process to generate final scenery texture $\mathbf{I}_s \in \mathbb{R}^{H \times W \times 4}$ is formulated as:
\begin{equation}
    \mathbf{I_s} = \mathcal{R} \left( \mathcal{G}_i \left(  \mathbf{M}_c, \mathbf{I}_b; \mathcal{C}_s \right)  \right).
\end{equation}
For foreground scenery that already contains an alpha channel in its template model, we directly reuse its alpha as $\mathbf{M}_c$ to ensure the correct structure.

\subsection{Multi-Modal Immersion Enhancement}
\label{ssec:method_immersion_enhance}

To further enhance immersion beyond static 3D visuals, we introduce agent-guided multi-modal enhancement in visual dynamics and sounds (see the right part of Fig.~\ref{fig:framework}).
\paragraph{Dynamic shader-based effects.}
The immersive enhancer analyzes the scenery component of the generated scene, and adds shader-based dynamic effects for natural elements such as flowing water, drifting clouds, and falling rain.  
These effects are implemented using customizable shader parameters, including procedural flow maps, noise-based motion textures, and screen-space animations, which bring liveliness to the scene while maintaining real-time performance (see supp. Sec. 1.3 for details).
\paragraph{Ambient sound synthesis.}
The immersive enhancer then synthesizes ambient sounds using a library of natural soundtracks tagged by content.  
Specifically, the agent analyzes the rendered panorama of the complete scene and retrieves suitable natural soundtracks (such as birds, winds, and water) from the library.
To support uninterrupted playback, we apply crossfading to seamlessly mix tracks for audio looping (see supp. Sec. 1.4 for details).

\section{Experiments}
\subsection{Implementation Details} 
We utilize Blender~\cite{blender} as our core scene modeling framework, integrating terrain texture projection, asset placement, scene rendering, and VR-ready scene export. 
We develop world modeling agents powered by GPT-4o, each configured with distinct system prompts (see supp. Sec.~1.2). The terrain library is primarily generated with Blender's A.N.T. Landscape add-on, resulting in a collection of about 10 initial templates for texturing. 
We build an asset library of about 50 foreground object templates, each modeled using alpha cards. The number of asset types when generating scenes ranges from 0 to 10, determined adaptively by the agents. The distance ranges for foreground and midground objects are configured as 2–10 meters and 20–50 meters, respectively.
The terrain-conditional diffusion model is fine-tuned from SDXL~\cite{podellsdxl} on 10K equirectangular terrain images collected from Unreal Engine (UE) scene rendering and Internet sources, with a learning rate of $1 \times 10^{-5}$ for 30K steps and batch size 4.
To further constrain the model when generating ERP images, we apply a circular padding to ensure continuity between the leftmost and rightmost edges of the panorama~\cite{zhang2024taming,feng2023diffusion360}.
For training the depth ControlNet~\cite{zhang2023adding}, we generate mixed depth maps using Depth-Anything V2~\cite{depth_anything_v2} and MiDaS~\cite{birkl2023midas} with perspective-to-panorama fusion, applying random scale and shift augmentations to improve robustness. We adapt VitMatte~\cite{yao2024vitmatte} with tile-based matting for high-resolution sky segmentation and PowerPaint~\cite{powerpaint} for sky outpainting. We leverage pre-trained diffusion models~\cite{layerdiffuse,powerpaint} for training-free RGBA texture synthesis. To balance realism and performance, we pre-bake high-resolution panoramic maps under global illumination as run-time unlit materials, enabling photorealism without the need for real-time lighting for VR applications (see supp. Sec.~1.6). All models are trained on a single NVIDIA A100-80G GPU. The entire generation pipeline takes about 10 minutes deployed on a single NVIDIA RTX 4090 GPU (see supp. Sec.~1.7).

\begin{figure*}[t!]
    \centering
    \includegraphics[width=1.0\linewidth, trim={0 0 0 0}, clip]{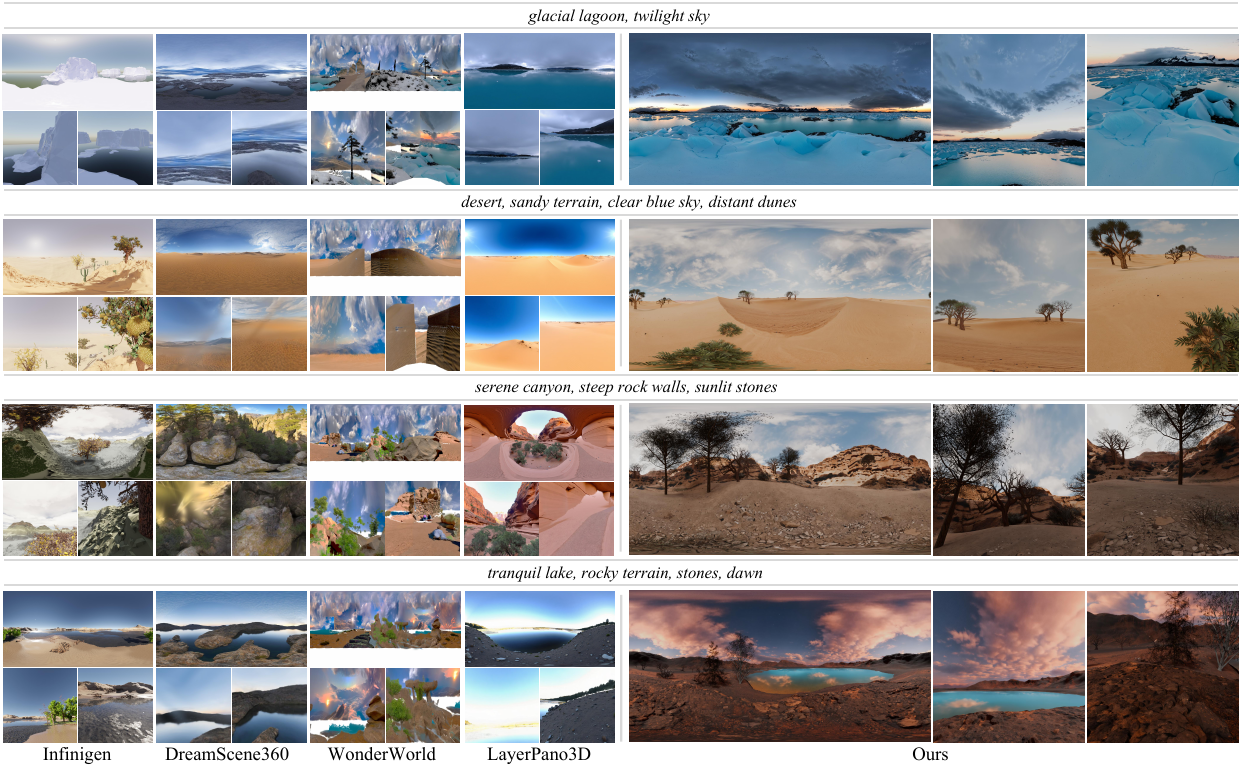}
    \caption{
    We compare our method with Infinigen~\cite{infinigen}, DreamScene360~\cite{dreamscene360}, WonderWorld~\cite{wonderworld} and LayerPano3D~\cite{yang2024layerpano3d} based on the generated 3D scenes using identical text prompts, visualizing both panoramic and perspective views of the generated scenes.
    }
    \label{fig:exp_comp}
\end{figure*}
\begin{figure*}[t!]
    \centering
    \includegraphics[width=1\linewidth, trim={0 0 0 0}, clip]{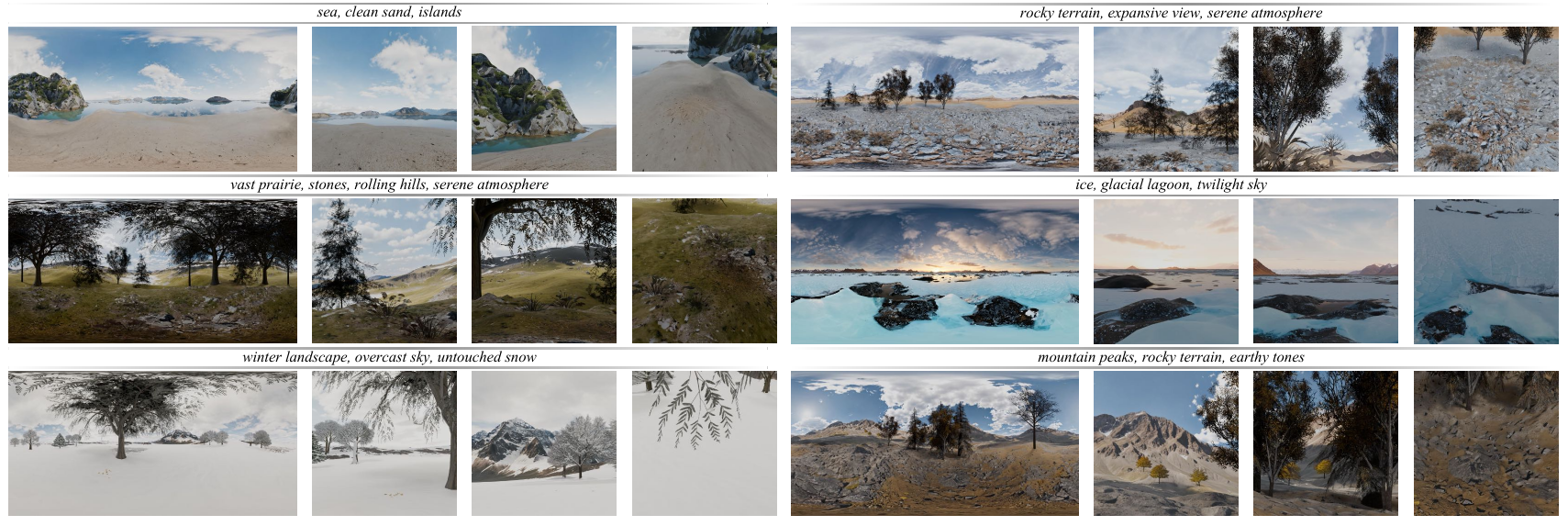}
    \caption{
    We present more examples of generated environments in panoramic and perspective views.
    }
    \label{fig:supp_gallery}
\end{figure*}

\subsection{Comparison on Scene Generation}
\label{ssec:exp_compare}

\paragraph{Baselines.} 
We compare our method with recent scene generation methods across different categories: 
(1) Infinigen~\cite{infinigen2023infinite}, which uses procedural generation with physics-based modeling;
(2) DreamScene360~\cite{dreamscene360}, which lifts panoramic images to 3D space;
(3) WonderWorld~\cite{wonderworld}, which generates scenes through perspective outpainting.
(4) LayerPano3D~\cite{yang2024layerpano3d}, similar to DreamScene360, but adopts a layered representation. 
For a fair comparison, we use Infinigen's scene configurations that match the same category with our generated scenes, adopt the same enhanced text prompts as our method for DreamScene360 and LayerPano3D, and use the cropped perspective images from our generated panorama as the image condition for WonderWorld.

\paragraph{Metrics.}
For comprehensive comparison with the above methods, we use metrics for evaluating both prompt-scene consistency and aesthetic quality, including CLIP similarity score (CLIP-Score)~\cite{radford2021learning}, aesthetic score (CLIP-Aesthetic)~\cite{aesthetic_predictor} and the VLM-based visual scorer Q-Align (QA-Quality)~\cite{wu2023qalign}.

\begin{figure*}[t!]
    \centering
    \includegraphics[width=1.0\linewidth, trim={0 0 0 0}, clip]{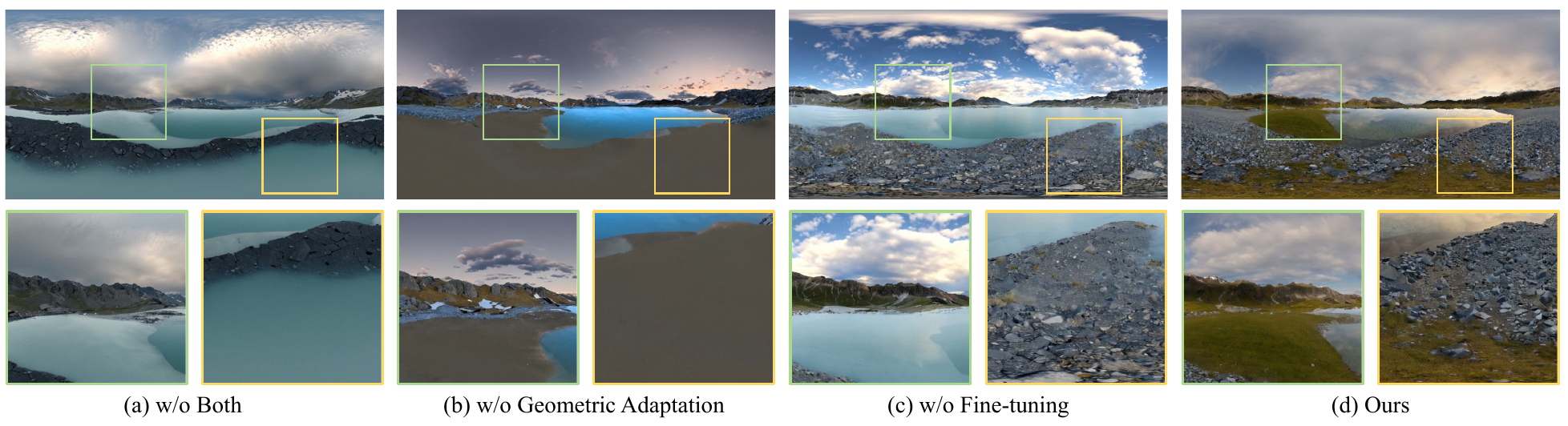}
    \caption{
    We analyze the geometric adaptation and fine-tuning of the conditioning network for terrain-conditioned texture generation.
    }
    \label{fig:exp_ablation_terrain_cond}
\end{figure*}
\begin{figure*}[t!]
    \centering
    \includegraphics[width=1.0\linewidth, trim={0 0 0 0}, clip]{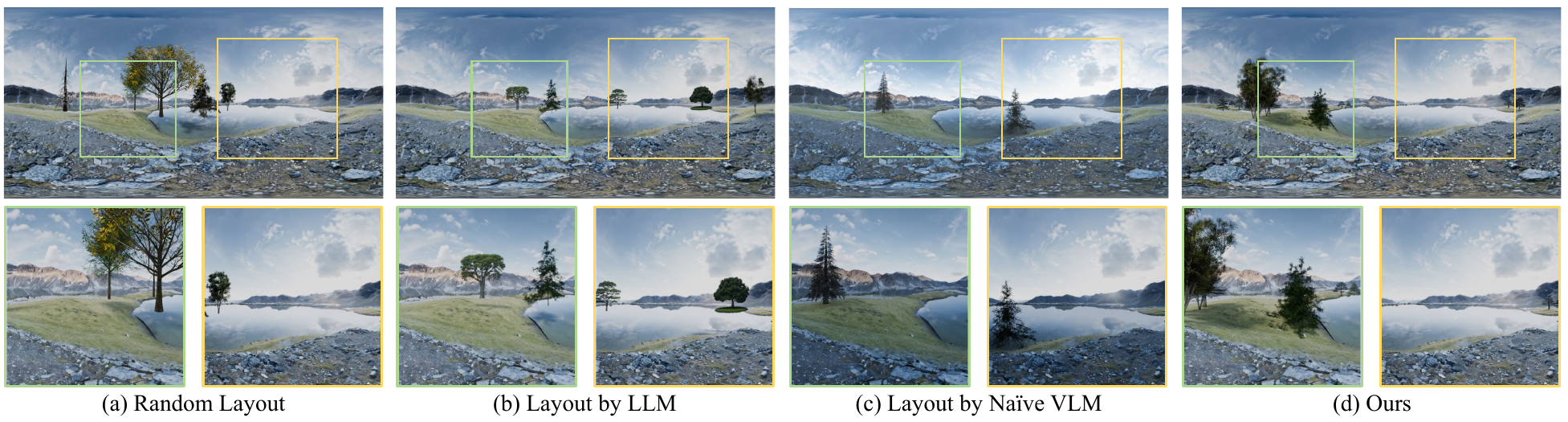}
    \caption{
    We compare our semantic grid-based analysis for asset layout with  different layout approaches.
    } 
    \label{fig:exp_ablation_layout}
\end{figure*}

\paragraph{Quantitative results.}

We present the quantitative comparison of our method with the baselines in Tab.~\ref{tab:quan}.
We generate 18 scenes for each method and render video sequences along circular camera trajectories, following the protocol of previous work~\cite{dreamscene360}. Evaluation metrics are reported as the average scores across all frames and scenes. The generation prompts encompass a wide spectrum of natural landscape settings (e.g., glaciers, beaches, and forests) to evaluate the generalizability across diverse environments. (as illustrated in Fig.~\ref{fig:exp_comp}).
As shown in Tab.~\ref{tab:quan}, our method outperforms all baselines in CLIP-Aesthetic score and QA-Quality, demonstrating the superior visual quality of our generated scenes.
For CLIP-Score, DreamScene360 and LayerPano3D also achieve competitive scores since they prioritize semantic alignment during training, while our method generates diverse textures(\eg, various geographic features instead of bare ground, see Fig.~\ref{fig:exp_comp}).

\begin{table}[t!]

\caption{
We perform quantitative comparison on the generated 3D scenes, and compare the complexity of representation (primitive count) and runtime performance (FPS) on VR devices.}
\resizebox{1.0\linewidth}{!}{
\tabcolsep 2pt
\begin{tabular}{lcccccc}
\toprule
\multicolumn{1}{c}{\multirow{2}{*}{Methods}} & \multicolumn{3}{c}{Quantitative Metrics} & \multicolumn{2}{c}{Complexity \& Perform.} \\ \cmidrule(lr){2-4} \cmidrule(lr){5-6}
\multicolumn{1}{c}{} & \multicolumn{1}{l}{CLIP-Score $\uparrow$} & \multicolumn{1}{l}{CLIP-Aesthetic $\uparrow$} & \multicolumn{1}{l}{QA-Quality $\uparrow$} & \multicolumn{1}{l}{Prim. Count $\downarrow$} & \multicolumn{1}{c}{FPS $\uparrow$} \\ 

\midrule
Infinigen     &   -         &   4.9546   &   3.0426  & \underline{1276k}   &  $\sim$7   \\ 

WonderWorld    &  27.0417   &   5.0116   &   2.6298   & 1632k   &  \underline{$\sim$14} \\ 

DreamScene360   &  \underline{29.3556}    &   4.8283   &   2.1446   & 2097k   &  $\sim$8 \\ 

LayerPano3D   &  \textbf{29.4633}    &   \underline{5.1513}   &   \underline{3.4812}   &  14577k   &  N/A  \\ 
Ours   &  28.8933   &   \textbf{5.4834}   &   \textbf{3.5445}  & \textbf{223k}    &  \textbf{$\sim$79}  \\ 
\bottomrule
\end{tabular}
}
\label{tab:quan}
\end{table}

\paragraph{Qualitative results.}
We present the qualitative comparisons in Fig.~\ref{fig:exp_comp}, evaluating the results based on several criteria: visual quality (absence of blurring or artifacts), diversity (the range of landscape features and scenery textures), and coherence (spatial consistency across the scene). Infinigen, relying primarily on limited procedural generators, produces scenes lacking diversity (e.g., monotonous ice, row 1) and coherence (e.g., contextually incompatible trees, row 4).
For DreamScene360, although it achieves consistent views with a panoramic lifting strategy, it lacks diverse scenery contents and also shows blurry artifacts (see the slanting floaters in the perspective views from the second and third row in Fig.~\ref{fig:exp_comp}) due to the instability of inpainting-based optimization and the limited resolution of 3D Gaussians.
For WonderWorld, since it relies on outpainting to generate a complete world, it cannot ensure view consistency across different views and results in fragmented scenes.
LayerPano3D produces aesthetic and consistent results with DiT-based panorama generator, but is prone to blurry artifacts and visible gaps at the layer boundary.
By contrast, our method builds up the world with hierarchical alpha-textured proxies while considering the 3D coherence with agent-guided modeling, preserving consistent quality across views and delivering immersive scenery content. We provide more examples of generated realistic scenes in Fig. \ref{fig:supp_gallery} and examples of scenes in a variety of styles and environments, demonstrating the generalizability of our approach.

\begin{figure*}[t!]
    \centering
    \includegraphics[width=1\linewidth, trim={0 0 0 0}, clip]{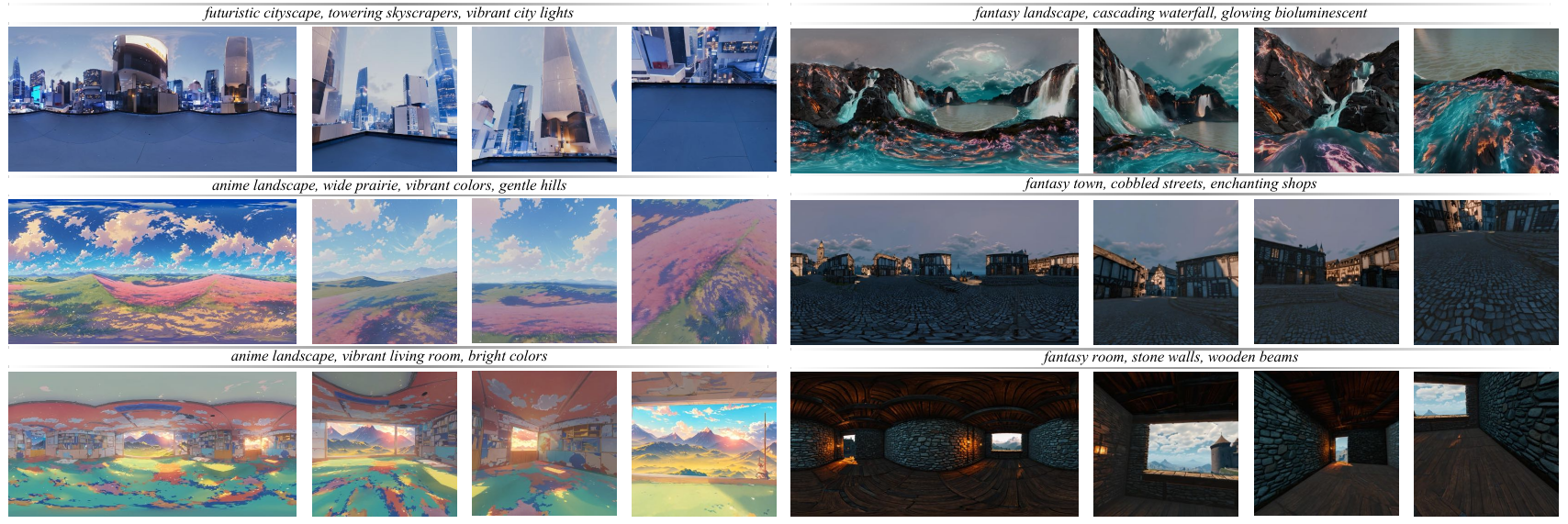}
    \caption{
     We present examples of generated scenes in various styles and urban or indoor environments beyond outdoor natural settings.
    }
    \label{fig:style_gallery}
\end{figure*}

\paragraph{User study.}
We conducted a user study involving 50 participants (33 of whom had expertise in 3D or computer graphics) to evaluate the 18 generated scenes. Using the PICO 4 Ultra VR headset, participants viewed randomized scene groups without time constraints and performed a forced-choice preference task across three aspects: Perceptual Quality (visual aesthetics and the absence of artifacts), Realism \& Coherence (environmental plausibility and spatial consistency), and Textual Alignment (adherence to the input prompt). LayerPano3D was excluded as its primitive count precluded VR rendering.  As shown in Tab.~\ref{tab:user}, our method was significantly preferred over baselines, demonstrating superior visual fidelity and alignment. This research was conducted in accordance with institutional policies, which did not require an ethical review. Informed consent was obtained from all participants before they attended the study.

\begin{table}[t!]
\caption{
We perform user studies on the generated 3D scenes.}
\resizebox{1.0\linewidth}{!}{
\tabcolsep 2pt
\begin{tabular}{lccc}
\toprule
Method                      & Perceptual Qual. $\uparrow$   & Realism \& Coherence $\uparrow$   & Textual Align. $\uparrow$ \\ 
\midrule
Infinigen                   &   7.12\%   &   5.83\%   &   6.04\%     \\ 
% \hline
WonderWorld                 &  13.46\%   &   7.50\%   &   11.49\%    \\ 
% \hline
DreamScene360               & \underline{24.01\%}    &   \underline{33.89\%}   &   \underline{38.22\%}    \\ 
% \hline
Ours                        &  \textbf{55.41\% }  &   \textbf{52.78\%}  &   \textbf{44.25\%}    \\ 
\bottomrule
\end{tabular}
}

\label{tab:user}
\end{table}

\paragraph{Complexity of representation and performance.}

We compare the complexity of scene representation and runtime performance on VR devices (Snapdragon XR2 Gen 2 platform). We calculate the average primitive counts and FPS of all scenes for each method.
As shown in Tab.~\ref{tab:quan}, methods using 3D Gaussians as representation (DreamScene360\cite{dreamscene360} and WonderWorld\cite{wonderworld}) either generally achieve only 8-14 FPS even with foveated rendering, or fail to launch on VR devices\cite{layerpano3d}. 
For Infinigen\cite{infinigen}, since it generates a detailed world with intricate procedural geometry and materials from generators, it remains computationally expensive for real-time rendering.
In contrast, our method maintains a compact representation while preserving scene quality, achieving an average FPS of \textbf{79+} on VR devices.

\subsection{Ablation Studies}
\label{ssec:exp_ablation}

\paragraph{Geometric adaptation.}
We first analyze the geometric adaptation for projected terrain depth and fine-tuning of the conditioning network in terrain-conditioned texturing (Sec.~\ref{ssec:method_base_world}). As shown in Tab.~\ref{tab: depth_abl}, we evaluate the QA-Quality for rendered sequences of 10 generated base world in different configurations. By ablating both strategies, the generated terrain texture fails to produce a plausible texture (water area on the bottom in Fig.~\ref{fig:exp_ablation_terrain_cond} (a)).
By enabling fine-tuning, the terrain texture precisely reflects the ground but with a monotonous appearance (see Fig.~\ref{fig:exp_ablation_terrain_cond} (b)).
By enabling geometric adaptation, the ground texture shows more detail (rocks on the bottom in Fig.~\ref{fig:exp_ablation_terrain_cond} (c)).
With all the strategies, we achieve terrain texture with fine-level details and realistic structure (see Fig.~\ref{fig:exp_ablation_terrain_cond} (d)). % 

\begin{table}[t!] 
\caption{Ablation study of the proposed geometric adaptation in terrain-conditioned texturing.}
\resizebox{1.0\linewidth}{!}{
\tabcolsep 8pt
\begin{tabular}{lcccc}
\toprule
          & w/o both  & w/o Geo. Ada.  & w/o Fine-tuning   & Ours \\ 
\midrule
QA-Quality $\uparrow$  & 3.6938    &  3.7630  &  3.7614    &  \textbf{3.9057}      \\ 
\bottomrule
\end{tabular}
}
\label{tab: depth_abl}
\end{table}

\paragraph{Semantic grid-based analysis.}
\label{ssec:exp_layout}
We then evaluate the efficacy of the proposed semantic grid-based analysis for the asset arranger (Sec.~\ref{ssec:method_asset_gen}).
Specifically, we compare our method with different strategies, including random layout generation, LLM-based generation that outputs object coordinates directly, and na\"ive VLM-based generator that receives unmodified base world images.
As shown in Fig.~\ref{fig:exp_ablation_layout}, the output of random layout incorrectly places trees on the lake (Fig.~\ref{fig:exp_ablation_layout} (a).
The layout generated by generic LLM and na\"ive VLM improves the coherence by providing compatible texture descriptions and plausible coordinates, but still suffers from inappropriate placements. By using semantic grid-based visual prompts as input for the VLM, our method generates a pleasant scene composition while addressing the placement issue.
As shown in Tab.~\ref{tab: layout_abl}, our layout generator achieves higher CLIP-Aesthetic scores comparing rendered panoramas from scenes generated by different layout generation strategies, demonstrating the efficacy of the semantic grid-based visual prompt for improving the placement quality.
\begin{table}[t!]
\caption{Ablation study of the semantic grid-based analysis.}
\resizebox{1.0\linewidth}{!}{
 \tabcolsep 8pt
\begin{tabular}{lcccc}
\toprule
Method          & Random layout  & LLM & Na\"ive VLM    & Ours \\ 
\midrule
CLIP-Aesthetic $\uparrow$   &    5.4963      &  5.5212   &  5.5350     &  \textbf{5.5739}      \\ 
\bottomrule
\end{tabular}
}
\label{tab: layout_abl}
\end{table}

\begin{figure}[t!]
    \centering
    \includegraphics[width=1\linewidth, trim={0 0 0 0}, clip]{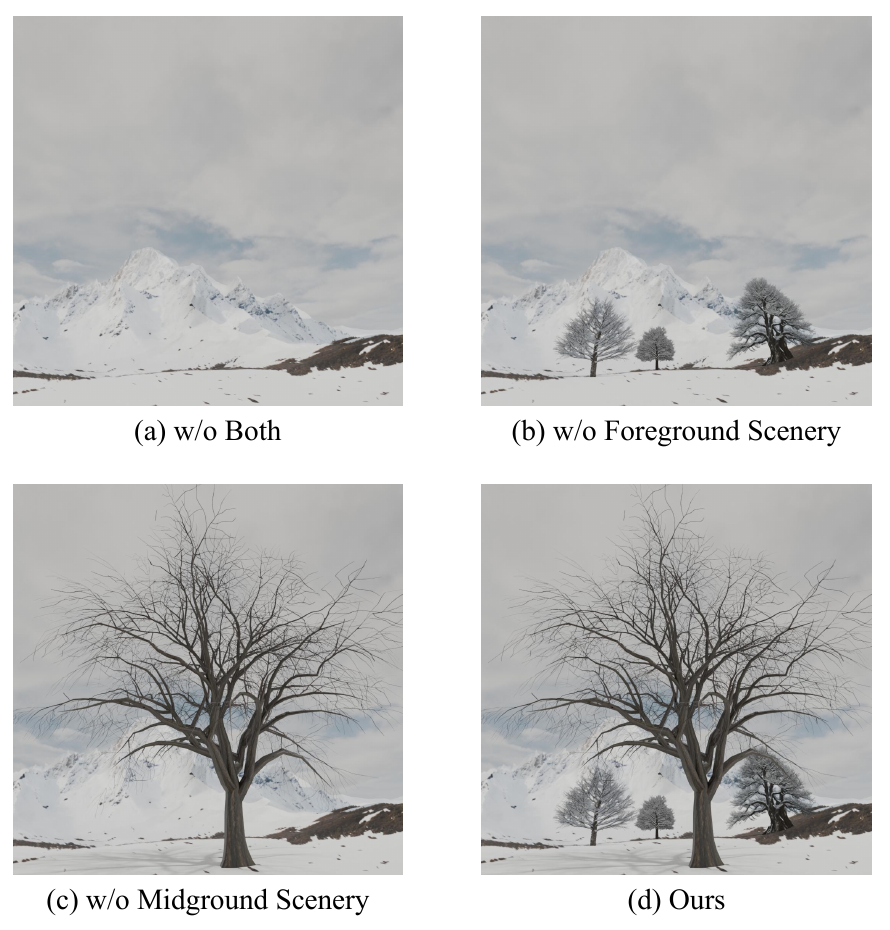}
    \caption{
    We visualize the contribution of different scenery by ablating proxy scenery of different types.
    }
    \label{fig:exp_ablation_lod_content}
\end{figure}

\begin{table}[t!]
% \vspace{2em}
\caption{
Ablation study on the aesthetic improvement of adding proxy scenery.
}
\resizebox{1.0\linewidth}{!}{
\tabcolsep 3pt
\begin{tabular}{lcccc}
\toprule
         & w/o Both scenery  & w/o Midground scenery  & w/o Foreground scenery   & Ours \\ 
\midrule
QA-Aesthetic $\uparrow$   &  2.1143     &  2.3287           &  2.3562             & \textbf{2.4408}   \\ 
CLIP-Aesthetic $\uparrow$   &  5.1540     &  5.3307           &  5.2453             & \textbf{5.3634}   \\ 
\bottomrule
\end{tabular}
}

\label{tab: asset_abl}
\end{table}

\paragraph{Aesthetic contribution with proxy scenery.}
We also investigate the aesthetic contribution when adding generated proxy scenery into the base world.
Specifically, we evaluate the QA-Aesthetic~\cite{wu2023qalign} and CLIP-Aesthetic score of 10 randomly selected generated scenes in absence of midground or foreground assets.
As shown in Tab.~\ref{tab: asset_abl} and Fig.~\ref{fig:exp_ablation_lod_content}, the added scenery significantly improves the visual quality by enriching the base world with diverse elements and improving the sense of depth.

\section{Discussion}
\subsection{Design Choices}
\begin{figure}[t!]
    \centering
    \includegraphics[width=1\linewidth, trim={0 0 0 0}, clip]{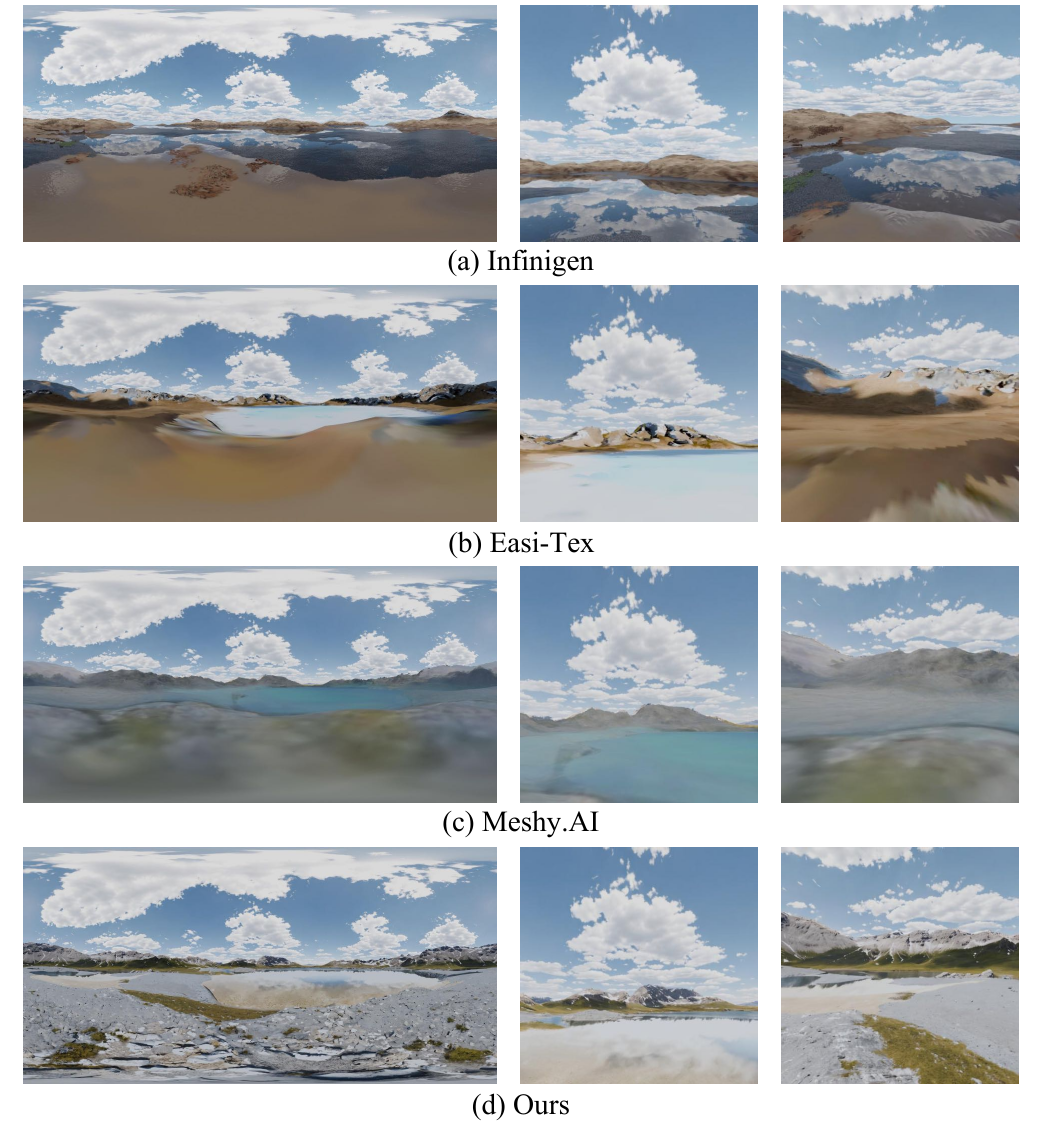}
    \caption{
    Our method outperforms other works on terrain texturing.
    }
    \label{fig:compare_texture}
\end{figure}

\paragraph{Panoramic terrain texturing.} We observe that existing methods often struggle to synthesize photorealistic terrain, typically producing overly smooth or blurry textures, such as the procedural generation framework Infinigen\cite{infinigen2023infinite}, generative texturing model Easi-Tex\cite{perla2024easitex} and commercial texturing tools Meshy.AI\cite{meshy}. As shown in Fig.\ref{fig:compare_texture} and Tab.\ref{tab:texturing_comparison}, our method achieves better texture quality benefiting from the proposed terrain-conditioned panoramic texturing and user-centric UV mapping.

\begin{table}[t!]
\caption{Quantitative Comparison on Terrain Texturing.}
\resizebox{1.0\linewidth}{!}{
\tabcolsep 15pt
\begin{tabular}{lccc}
\toprule
Method & Type & CLIP-Aesthetic $\uparrow$ & QA-Quality $\uparrow$ \\ 
\midrule
Infinigen & Procedural & 5.2937 & 2.9091 \\
Easi-Tex  & Generative & 4.8599 & 2.9242 \\
Meshy.AI  & Commercial & 4.8750 & 3.2685 \\
Ours & Generative & \textbf{5.4317} & \textbf{3.4860} \\
\bottomrule
\end{tabular}
}
\label{tab:texturing_comparison}
\end{table}

\begin{figure}[t!]
    \centering
    \includegraphics[width=1\linewidth, trim={0 0 0 0}, clip]{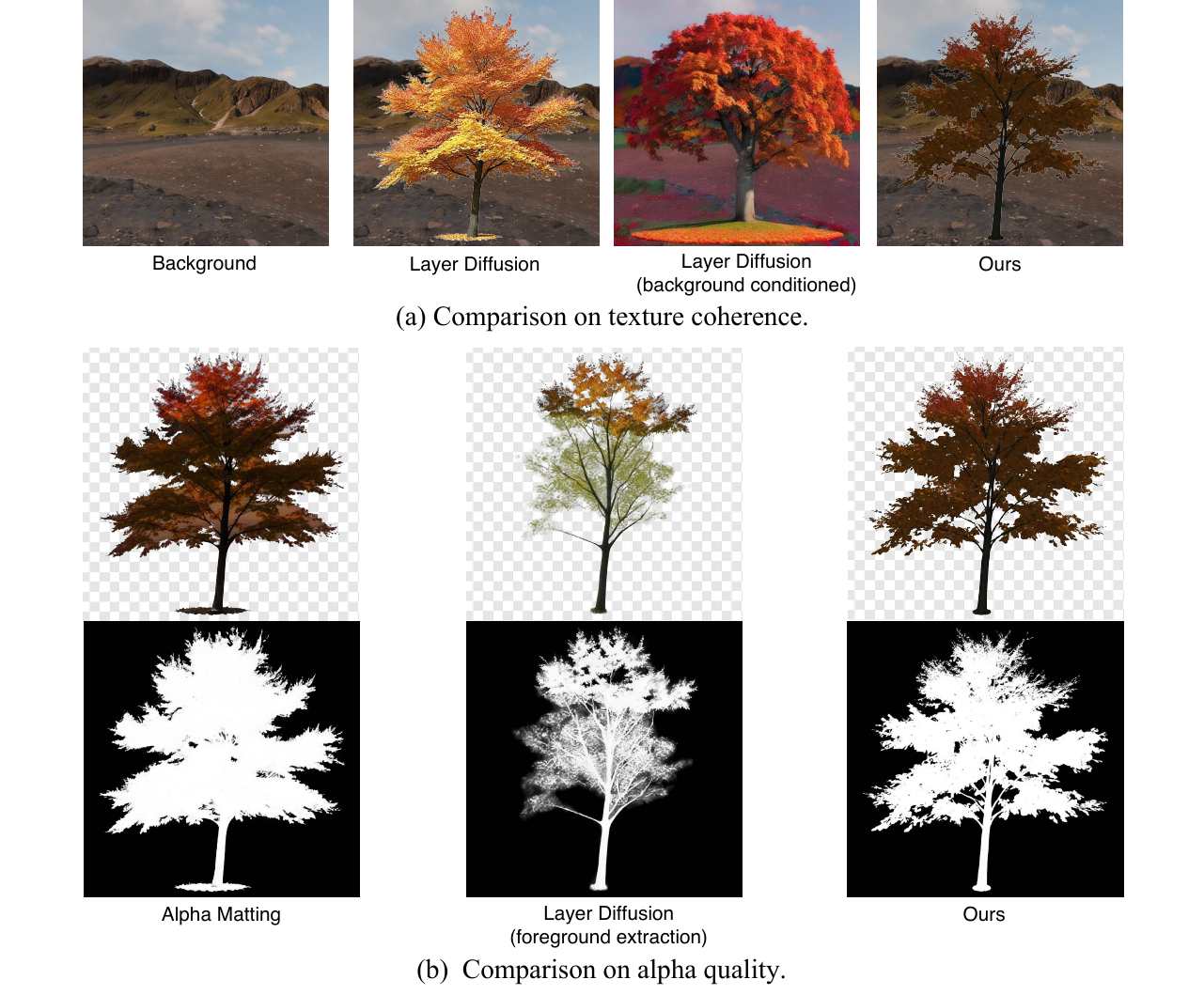}
    \caption{
    Our RGBA texture synthesis achieves better coherence and quality comparing with the baseline.
    }
    \label{fig:alpha}
\end{figure}

\paragraph{Context-aware asset texturing.}
While off-the-shelf generative models can produce asset textures, achieving precise and coherent integration of these assets into diverse backgrounds remains challenging. 
We visualize the texture generated from the proposed context-aware RGBA texture synthesis and the original layered diffusion model~\cite{layerdiffuse}. As shown in Fig.~\ref{fig:alpha}, our generated assets demonstrate greater consistency with the background and alpha quality, attributed to our disentangled generation of color and the alpha channel. 

\begin{figure}[t!]
    \centering
    \includegraphics[width=1\linewidth, trim={0 0 0 0}, clip]{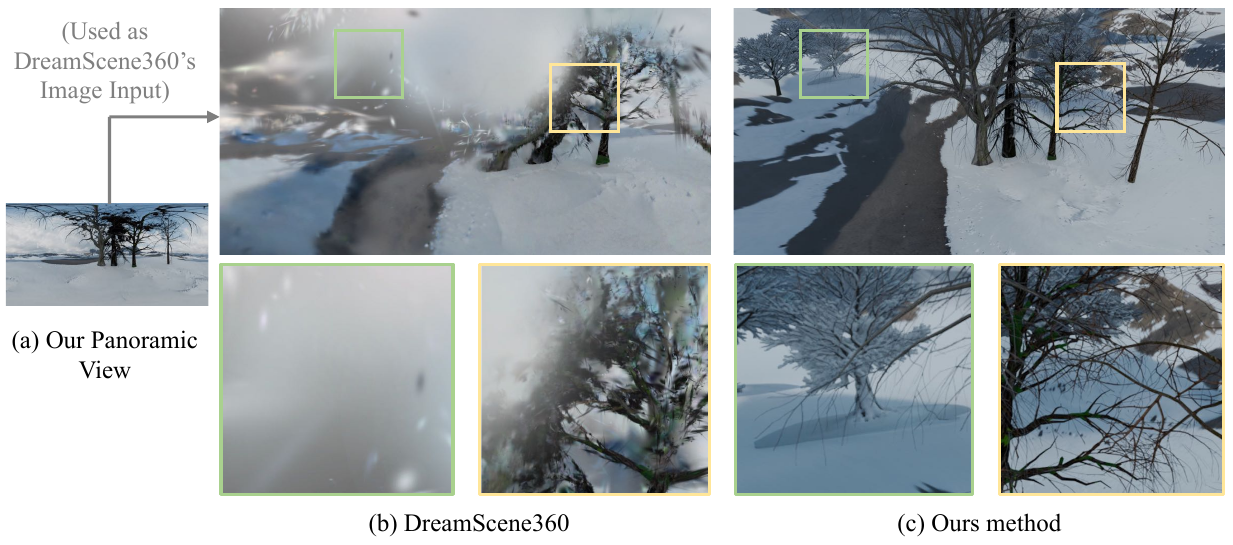}
    \caption{
    Our method produces scenes with higher visual fidelity compared to methods based on 3D Gaussians.
    }
    \label{fig:compare_ds360}
\end{figure}

\paragraph{Proxy-based representation.}
To further evaluate the fidelity of our proxy-based representation relative to 3D Gaussians, we adapt DreamScene360~\cite{dreamscene360} to generate scenes on the same panorama image from our method.
As shown in Fig.\ref{fig:compare_ds360}, the generated 3D Gaussians scenes exhibit notable artifacts and reduced visual fidelity, which is further reflected in lower metrics as reported in Tab.~\ref{tab:ds360_img}, indicating its limitations in representing high-quality scenes compared to our approach.
\begin{table}[t!]
% \vspace{2em}
\caption{Comparison with 3D Gaussians using the same panorama.}
\resizebox{1.0\linewidth}{!}{
\tabcolsep 15pt
\begin{tabular}{llll}
\toprule
Method          & CLIP-Score $\uparrow$   & CLIP-Aesthetic $\uparrow$   & QA-Quality $\uparrow$ \\ 
\midrule
DreamScene360    &  28.2111    &   4.8748   &   2.2975         \\
Ours             &  \textbf{28.7806}   &   \textbf{5.3289}   &   \textbf{3.2066}          \\ 
\bottomrule
\end{tabular}
}
\label{tab:ds360_img}
\end{table}

\subsection{Limitations and Future Work}
First, our method focuses on outdoor scenes rather than indoor scenes with detailed furniture. 
Second, the scenes are currently restricted to a limited exploration range (typically around $50\,m^2$) due to the fixed hierarchy of generation levels relative to viewing distance.
This could be addressed by incorporating novel view synthesis or video generation techniques~\cite{gu2025das} to create extensible scenes in future work.
Third, our approach relies on pre-built templates for foreground assets, which could be enhanced by integrating procedural generators~\cite{speedtree} to enable the creation of more diverse templates.

\section{Conclusions}
We have presented ImmerseGen, a novel framework for generating photorealistic 3D environments from lightweight geometric proxies tailored for immersive experiences. The proposed generative terrain-conditioned texturing and alpha-textured asset synthesis eliminate the need for dense modeling to create diverse scenes. VLM-based agents orchestrate the entire pipeline—ranging from asset selection, design, and arrangement to multi-modal immersion enhancement. Our method creates coherent worlds while maintaining real-time rendering on mobile platforms.

\bibliographystyle{abbrv-doi-hyperref}
\bibliography{main}

\end{document}